# Deciphering the complex intermediate role of health coverage through insurance in the context of well-being by network analysis

Myriam Patricia Cifuentes[1]
*College of Public Health and Department of Biomedical Informatics*
*The Ohio State University*

Soledad A. Fernandez
*Center for Biostatistics, Department of Biomedical Informatics*
*The Ohio State University*

**ABSTRACT**

Recent initiatives that overstate health insurance coverage for well-being conflict with the recognized antagonistic facts identified by the determinants of health that identify health care as an intermediate factor. By using a network of controlled interdependences among multiple social resources including health insurance, which we reconstructed from survey data of the U.S. and Bayesian networks structure learning algorithms, we examined why health insurance through coverage, which in most countries is the access gate to health care, is just an intermediate factor of well-being. We used social network analysis methods to explore the complex relationships involved at general, specific and particular levels of the model. All levels provide evidence that the intermediate role of health insurance relies in a strong relationship to income and reproduces its unfair distribution. Some signals about the most efficient type of health coverage emerged in our analyses.

**INTRODUCTION**

The third goal of the 2030 Agenda for Sustainable Development of United Nations (U.N.), "Ensure healthy lives and promote well-being for all at all ages,"[1,2] includes the time-honored idea of paralleling health with wellbeing.[3] This comprehensive commitment broadens the spectrum of health by including nine targets and four related topics that reprise the three former Millennium Development Goals[4] and add aspects covering health results and processes. Furthermore, by declaring that all 16 sustainable development goals and targets are integrated and indivisible,[1] the 2030 Agenda reaffirms health relation to well-being.

When individuals face disease, loss of earnings due to inability to work concurrent with climbing health care expenses devastates people's and household's economy. This scenario justifies that the U.N. and the World Health Organization emphasize the target of universal health coverage. However, presenting health care coverage as the way to promote health and well-being and extend life expectancy[1] and as "the single most powerful concept that public health has to offer,"[5] narrows the comprehensive commitment and contradicts well known evidence about the health care as an intermediate determinant factor of health.[6]

---

[1] cifuentesgarcia.1@osu.edu, mpcifuentesg@unal.edu.co

By financing people's health care expenses, insurance is an acknowledged way to accomplish health coverage. Simultaneously, health insurance allows for financing the health economic sector, which delivers these usually expensive services and goods mainly circumscribed to intervene disease.[7] However, the high costs of this narrow operationalization of health impugn efficiency, effectivity and contribution of health insurance to attain people's well-being. Therefore, different authors and initiatives such as the 'health and wealth' debate in Europe have revisited and questioned these complex relationships.[5,8–10]

Based on a quantitative systemic approach of a model that mapped all the statistical conditional dependences among variables available about resources with social value related to well-being, including health insurance, our recent results confirm the intermediate role of health insurance in well-being.[11] To explain why health insurance through coverage is just an intermediate factor of well-being, we focused on the health insurance subnetwork in our model to implement specific social network analysis methods.

Despite the ongoing implementation of the Patient Protection and Affordable Care Act (PPACA-Obamacare), we used the U.S. case because it provides a scenario where multiple health coverage schemes coexist.[12] Such variety allows for comparison useful for several implementations of health coverage by other countries.

**METHODS**

The model we used for our analysis was a complex network mapping all the statistical conditional independencies among all the variables available in the Integrated Public Use Microdata Series (IPUMS-USA) version of the 2013 American Community Survey (ACS).[13] This survey is a source of information for federal and state decision making about welfare investments aimed to improve population access to social resources.[14] The process to develop the model involved data preprocessing, network reconstruction by structure learning algorithms of Bayesian networks (also known as belief or causal networks) to control for spurious associations, and validation by data science methods.[11]

In the model each node corresponds to a variable, and a group of nodes belongs to a social resource according to the topics in the ACS (Figure 1).

According to the data dictionary of IPUMS-USA following the CB guidelines, the group describing the social resource of health insurance included 11 variables: A constructed variable named as 'any health insurance' that comprised private and public insurance; 'Public insurance' variable that contained insurance through Medicare, Medicaid and Veterans Administration (VA), but does not include the Indian health services variable; and 'Private insurance' variable that contained insurance through employer, plans purchased by individuals, and Tricare or other military health care.

As family relationships support health insurance eligibility of dependents, by using ACS variables such as age, sex, marital status, parents' location in the household, and number of children, IPUMS-USA developed a specific variable describing seven types of affiliations to health insurance units (HIU).[15]

After discarding significant redundancy associated with constructed variables by running collinearity tests at preprocessing, we included in the model variables of 'any health insurance', 'public insurance', 'private insurance' and 'HIU' as the CB and IPUMS-USA issued them to provide insights relevant to health insurance analysis.[15]

Different network science methods allowed us to navigate global and local scales of the model by general, specific and particular levels of analysis.

Specific analysis of the U.S. case zooms into the subnetwork of health insurance to introduce explanations of general results. On the same scale, particular analysis focuses on the details of links between single variables of health insurance and other social resources.

Network science methods identify connectivity patterns by visualization techniques, calculation of indices and community detection. For centrality indices of degree, closeness, betweenness, and weights as hubs or authorities[16,17] we calculated estimators such as means, percentiles (Pc) and z-scores (z). Functional forms of indices' distributions were the basis of known statistical tests to identify salient nodes and relationships. We examined pairwise relations and other groups of linked variables for interpreting results. Groups of three-linked nodes (i.e. triads), comprising health insurance and different social resources allowed for brokerage analysis to approach mediation roles such as coordinator, itinerant broker, gatekeeper and liaison, described in social network analysis.[17] We detected emerging groups of variables by the Louvain method.[17–19] These methods were available in Pajek software.[20] We omitted analysis of links' direction.

**RESULTS**

**1.     General analysis results**

General analysis presents the quantitative results supporting the intermediate role of health insurance in the global landscape of all observed social resources of well-being, each constituting a subnetwork (Figure 1). Subnetwork of health insurance connected with 15 (83.3%) subnetworks of other social resources by 29 shrunken links that consolidate all the links between nodes in each group, accounting for 15.67% of the entire network.

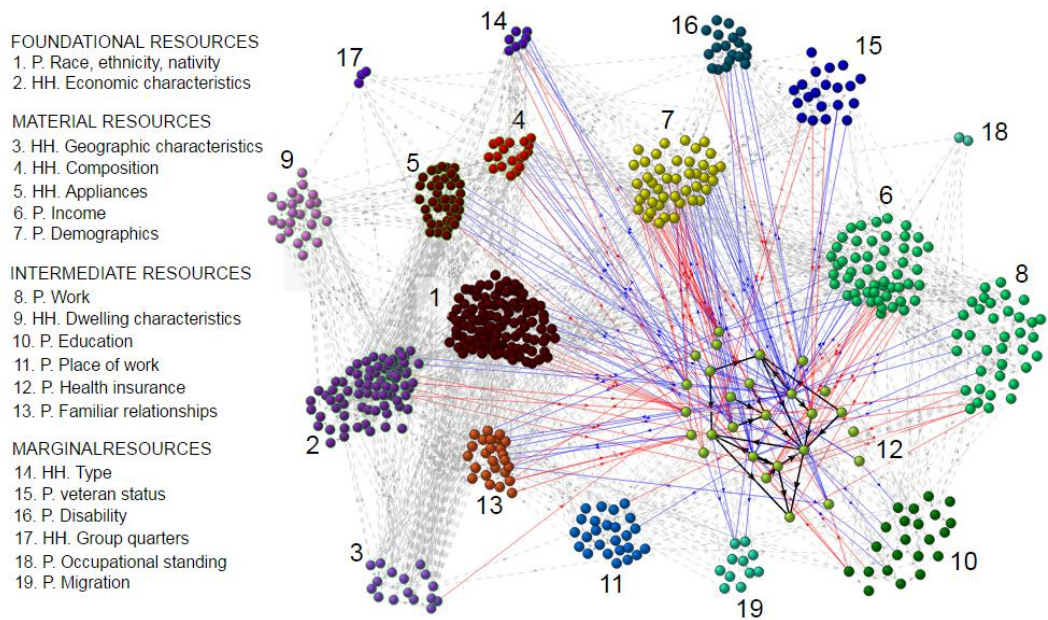

**Figure 1.** Network of subnetworks of social resources for well-being. The legend shows the ranking of social resources in four layers resulting from hierarchical clustering. Health insurance subnetwork (12) has black links among its own variables. Links between variables of health insurance and other social resources are red if incoming, and blue if outgoing.

According to centrality indices of each subnetwork, health insurance ranked sixth by degree (general, in and out-degree) and betweenness, and seventh by closeness centrality. Additionally, the permutation obtained from hierarchical clustering, showed health insurance subnetwork ranking twelfth. Hierarchical clustering also located health insurance at a third level of four, besides other service-based social resources, such as education, dwelling, place of work and familiar relationships. The four layers are presented in the legend of figure 1.[11]

By analyzing the distribution of weights of links between health insurance and the other social resources, we found significant bidirectional relationships with 'person demographics' and 'person income'. The relationship with 'culture', which included variables of race, ethnicity, nativity, residency continuity and language, was significant just for the outgoing link weights (Table 1 summarizes p-values, percentiles and z-scores for significant relationships).

Table 1. Summary of p-values (p) percentiles (Pc) and z-scores (z) for different indices of relevant relationships between nodes

| Rank | Node | Index | p-value |
|---|---|---|---|
| General analysis | | | |
| 1 | Person demographics | In-link weights | $p=4.25 \times 10^{-9}$ |
| | | Out-link weights | $p=5.43 \times 10^{-9}$ |
| 2 | Person income | In-link weights | $p=1.03 \times 10^{-4}$ |
| | | Out-link weights | $p=8.01 \times 10^{-4}$ |
| 3 | Culture | Out-link weights | $p=8.38 \times 10^{-7}$ |
| | | Out/in weight ratio | $p=1.22 \times 10^{-3}$ |
| Particular analysis | | | |
| 1 | No insurance through Medicare | Degree=36 | Pc 99.5, z=5.6 |
| | | Closeness | Pc=99.1 z=1.5 |
| | | Betweenness | Pc=97.3, z=3.3 |
| | | Hub weight | Pc=99.5, z=7.1 |
| | | Itinerant broker (31 triads) | Pc=99.1 z=7.6 |
| | | Liaison (152 triads) | Pc=99.8 z=12.3 |
| | | Gatekeeper index (14 triads) | Pc=95.0 z=1.8 |
| 2 | 'insurance through Medicare' | Degree=18, | Pc 96.7, z=2.15 |
| | | Closeness | Pc=97.1 z=1.0 |
| | | Hub weight | Pc=98.6 z=3.6 |
| 4 | without private insurance | Degree=15 | Pc=93.0 z=1.6 |
| | | Betweenness | Pc=99.8 z=5.6 |
| | | Representative | Pc=99.1 z=3.6 |
| | | Liaison | Pc=93.0 z=0.9 |
| | | Coordinator | Pc=92.5 z=1.0 |
| 3 | 'Single adults to self' | Degree=16 | Pc=93.8, z=1.8 |
| | | Closeness | Pc=94.7 |
| | | Betweenness | Pc=92.0 |
| | | Hub weight | Pc=93.5 |
| | | Authority weight | Pc= 92.1 |
| | | Itinerant broker | Pc=98.0 z=2.8 |
| | | Liaison | Pc=98.0 z=2.7 |
| 5 | 'Reference person to self' | Degree=13 | Pc=89.3 z=1.2 |
| | | Authority | Pc=91.4 z=1.7 |
| | | Itinerant broker | Pc=96.7 z=1.76 |
| | | Liaison | Pc=97.3 z=2.1 |

## 2. Specific analysis results

By focusing on health insurance subnetwork, we identified that not all variables were connected (Figure 2). The subnetwork included just 15 of 27 variables (55.5%), connected by 20 links. However, links are selective just over public types of insurance (Tricare with VA, and Medicare and Medicaid) and no insurance. Conversely, unlinked variables denote significant stand-alone types of insurance, such as private health coverage, insurance through employer, and Indian health services. Lack of connections for 'any health insurance' variable discarded the preeminence of a specific type of insurance. Almost all HUI rules were unconnected also.

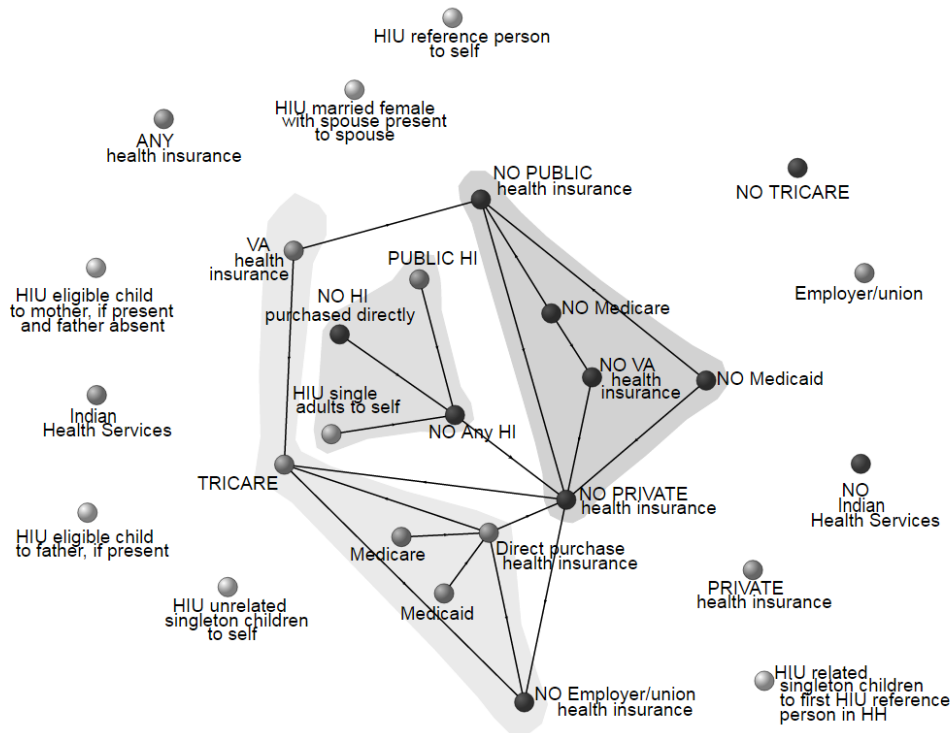

Figure 2. Subnetwork of health insurance variables (represented by nodes). Variables about type of insurance are mid-grey, no insurance variables are dark gray and eligibility rules are light gray. Shades of gray highlight three communities of nodes obtained by the Louvain method.

The locally most-connected variable of 'no private insurance' (degree Pc=93.7 and z =2.41, p=1.10x10$^{-7}$) carried the general meaning of the subnetwork, and marked the convergence of three communities identified by the Louvain method (Figure 2): The top right community gathered variables of no public insurance that were linked through 'no private insurance' to 'direct purchase'. The middle left community showed single adults as a profile significantly linked to 'no any health insurance'. The bottom left community signaled the coexistence of direct purchase of insurance with public insurance through Medicare, Medicaid, Tricare and no insurance through employer, as well as showing the overlap between Tricare and VA.

### 3. Particular analysis results

Links between single variables of health insurance and all remaining social resources accounted for 8.85% of the network connections (Blue and red links in Figure 1). We explored these links, emphasizing those belonging to the most relevant relationships of health insurance with demographics, culture and income (Table 2. Supplemental material available by request includes links with the remaining subnetworks).

Forty-five pairwise links constituted the relationship between health insurance and demographic characteristics. As demographics variables such as age, sex, and relation to household head were the basis to build HUI rules,[15] the links connecting these variables accounted for 62.22% of the relationship.

The remaining 37.78% links revealed people's profiles related to insurance types, such as that of 'no any health insurance' significantly associated to institutional inmates. Insurance through Medicare had significant associations with 'female sex' and with individuals from 15-64 years old. The complement variable of 'no health insurance through Medicare' had significant links with individuals between 0-4 and beyond 64 years old. This last disclose a relevant partial coverage of a target population by the Medicare program.

The relationship between income and health insurance had 24 links, 70% at the expense of insurance types. Variables of personal income included different sources of individual and family income, such as earnings, wages, interests, rentals, retirement income, social security income, and poverty level (PL). PL variable includes variables of family structure, income, and the official Federal Poverty Level (FPL) thresholds,[21] which define eligibility for diverse welfare assistance, including some types of public health insurance[22,23]. As the states differently implement federal thresholds, we automatically defined a general version of thresholds by applying clustering techniques to the observed distribution of the variable. We used the same method to define thresholds for income variables measured in continuous scale.

The variables of 'no any health insurance' and 'insurance through Medicaid' had significant links with the lowest PL from 1-147%, which includes FPL thresholds for Medicaid eligibility (133% before and 138% after PPACA), and with PL between 147-310%, which includes thresholds for assistance on out of pocket cost of silver plans (FPL 100-250% after PPACA) and for qualifying to premium tax credits (FPL 100-400%)[23]. Additionally, Medicaid also had significant links with variables registering neither incomes from welfare nor supplementary security.

Insurance through Medicare significantly linked to the highest retirement income, the lowest total personal incomes, and no income from social security. The complement of 'no insurance through Medicare' had significant links with the lowest retirement income, middle and highest social security income, and the full rank of supplementary security income. With no links to insurance or not through Medicare, middle retirement income linked to 'no private insurance'.

Table 2. Inventory of links between variables of health insurance and variables of the remaining social resources available in the 2013 American Community Survey (ACS)

| Relationship between health insurance and Person's demographics | | |
|---|---|---|
| Health insurance variable/node | link | Variable in other subnetwork |
| No any health insurance | ← | Relationship to household head: Institutional inmates |
| No Health insurance through Medicare | ← | Age 75-84 years |
| | ← | Age 85-Highest |
| | ← | Age 65-74 years |
| | → | Age Less than 1 year |
| | → | Age 1 year |
| | → | Age 2-4 years |
| | → | Divorced in the past year |
| | → | Married thrice (or more) |
| | → | Relationship to household head: Child |
| | → | Widowed in the past year |
| Health insurance through Medicare | → | Age 15-24 years |
| | → | Age 25-44 years |
| | → | Age 45-64 years |
| | → | Marital status: Widowed |
| | → | Sex: Female |
| Has health insurance purchased directly | ← | Age 45-64 years |
| HIU reference person to self | ← | No Divorced in the past year |
| | ← | Marital status: Married, spouse absent |
| | ← | Relationship to household head: Child |
| | ← | Relationship to household head: Parent-in-Law |
| | ← | Relationship to household head: Parent |
| | ← | Sex: Male |
| | → | Marital status: Married, spouse present |
| HIU married female with spouse present to spouse | ← | Marital status: Married, spouse present |
| | ← | Sex: Male |
| HIU eligible child to father, if present | ← | Relationship to household head: Child |
| HIU Eligible child to mother, if present and father absent | ← | Age 15-24 years |
| | ← | Relationship to household head: Sibling |
| | → | Relationship to household head: Child |
| HIU Single adults to self | ← | Age 15-24 years |
| | ← | Marital status: Married, spouse absent |
| | ← | Sex: Female |
| | ← | No Widowed in the past year |
| | → | Marital status: Married, spouse present |
| | → | Relationship to household head: Partner, friend, visitor |
| | → | Relationship to household head: Other relatives |
| | → | Relationship to household head: Grandchild |
| HIU related singleton children to first HIU reference person in HH | ← | Relationship to household head: Sibling |
| | → | Relationship to household head: Child-in-law |
| | → | Relationship to household head: Child |
| | → | Relationship to household head: Other relatives |
| | → | Relationship to household head: Grandchild |
| HIU unrelated singleton children to self | ← | Relationship to household head: Other non-relatives |
| | ← | Relationship to household head: Partner, friend, visitor |
| Relationship between health insurance and Person's income | | |
| No any health insurance | ← | Poverty status 1-147% PL |
| | ← | Poverty status 148-310% PL |
| Health insurance through Medicaid | ← | No Supplementary Security Income |
| | ← | No Welfare |
| | ← | Poverty status 1-147% PL |
| | ← | Poverty status 148-310% PL |
| No health insurance through Medicare | ← | Supplementary Security Income $10-5,300 |
| | ← | Supplementary Security Income $5,400-11,900 |
| | ← | Supplementary Security Income $12,000-30,000 |
| | → | Retirement income up to $22,400 |
| | → | Social Security income $98,00-18,300 |
| | → | Social Security income 18,400-50,000 |
| Health insurance through Medicare | → | Retirement income $76,000-178,000 |
| | → | Social Security income Zero |
| | → | Total personal income $1-48,890 |
| No health insurance through VA | → | Other income up to $21,800 |
| Without private health insurance coverage | ← | Retirement income $22,500-75,000 |
| | ← | Poverty status 1-147% PL |
| | ← | Poverty status 501% limit PL |
| No health insurance through employer/union | ← | Total personal income $48,900-176,300 |
| Has health insurance purchased directly | ← | Interest, dividend, and rental income Zero |
| HIU reference person to self | → | Other income up to $21,800 |
| | → | Wage and salary income $86,000-240,000 |
| HIU married female with spouse present to spouse | → | Total personal income $176,400-128,1000 |
| Relationship between health insurance and Person's race ethnicity an nativity | | |
| With any health insurance coverage | → | Not a US. citizen |
| No health insurance through Medicaid | → | Birthplace: Nebraska |
| No Health insurance through Medicare | ← | Number of major race groups 4 race groups |
| | → | Hispanic origin Mexican |
| Has insurance through Indian Health Service | ← | Birthplace: Alaska |
| | ← | Birthplace: Montana |
| | ← | Birthplace: Arizona |
| | ← | Birthplace: New Mexico |
| | ← | Birthplace: Oklahoma |
| | → | Race: American Indian or Alaska Native |
| Without private health insurance coverage | → | Does not speak English |
| | → | Speaks English very well |
| | → | Speaks only English |

Other significant links were those between health insurance through employer and middle total personal income, between no private insurance and PL beyond the upper limit, and between HIU reference person with middle high wages and other sources of income.

The relationships of health insurance with culture included 13 links, all between insurance types and variables of race, ethnicity, citizenship, time living in the US and birthplace, which broadly approach culture in terms of sense of belonging. The most connected health insurance variable was 'Indian Health Services,' that significantly linked to American Indians or Alaska Natives races and birthplaces of Alaska, Montana, Arizona, New Mexico and Oklahoma.

Significant links of no insurance through Medicare with multicultural households composed of four race groups, and with Mexican origin contrast with no private health insurance, equally linked to variables describing 'just English speakers' and 'no English speakers.'

Emerging results by looking at single links allowed us to identify particular roles of health insurance variables in the entire network. The most central variable was 'no insurance through Medicare' with high indices of degree, closeness, betweenness and hub weight. High indices as itinerant broker and liaison signaled its mediating role within and between social resources different to health insurance. According to its high gatekeeper index, it also mediated between social variables and the health insurance subnetwork, uncovering diverse population profiles without health insurance, such of children, elders with low retirement incomes, people being covered by social security, some regularly out of the working force, and some with recent family hardship.

The complement of 'insurance through Medicare' was the next highly-central variable with high degree, closeness and hub weight. However, it had no any mediation role in terms of other social resources. Besides significant links with young people, insurance through Medicare also included individuals with lower total income, and those who do not receive social security income, contrasting with retirees with high retirement income. This variable links with both people available to work and not.

'Without private health insurance' was the next health insurance variable ranking 4$^{th}$ according to high degree and the highest betweenness. Its mediator roles were as representative, as liaison and as coordinator. No private health insurance linked to both extremes of poverty levels, to English speakers and non-speakers, and selectively connected 10$^{th}$ and 11$^{th}$ grades of educational attainment.

Two HIU profiles ranked third and fifth. Ranking third, 'Single adults to self' had high degree, closeness and betweenness, hub and authority weights, none of which had high z-scores. Its role of mediation was as itinerant broker and as liaison. In addition to its significant link to 'no any health insurance', the profile of independent singles had significant associations with female sex and male and female householders living alone. Besides links with married 'people with spouse present' or 'absent', it also significantly linked to same sex married couples.

'Reference person to self' ranked fifth with a non-highly significant degree and a mild role as authority. However, it had more relevant roles as itinerant broker and liaison. The profile of reference person significantly linked to the male sex and married persons. However, it also linked to households headed by both male and females with absent spouse.

**DISCUSSION**

By social network science methods, we explored a model mapping multiple controlled dependencies among social resources for well-being, including health insurance. According to network indices, the ranking of a health insurance subnetwork behind other social resources for well-being confirmed its intermediate role. Even health insurance connects with all the best ranked social resources, its main relationships were with persons' income, demographic characteristics and culture.

These relationships relied on links with two classes of variables: the different types of insurance in the U.S., and the family-based eligibility profiles for health coverage.

Independently of whether race-specific health services recognize or segregate cultural differences, intuitive links of race and place of birth with Indian Health Services narrow the relationship of health insurance with culture. Regarding the relationship between health insurance and demographics, the majority of links were also intuitive because they were due to profiles of eligibility. Otherwise, dominant links between types of coverage and economic variables that explain the relationship of health insurance and persons' income are relevant and meaningful to explain why health coverage through insurance just plays an intermediate role in well-being.

The emergent subnetwork of health insurance, which include the links among multiple insurance types supports that sources of health coverage coexist.[24] However, the subnetwork revealed this coexistence just selectively affect public and lack of insurance, that signal a respectively incomplete or absent coverage of health needs. Furthermore, the links of public and lack of insurance with direct purchase of health plans, as if this were the only possible solution to get coverage, indicate that access to income unavoidably mediates access to a satisfactory health care.

In particular analysis, the links of Medicaid, as the only explicit income-focused public program, confirm it covers the target poorest populations. However, links of lower poverty levels with no any health insurance revealed the still significant lack of coverage. The significant link with the highest retirement income revealed the different pattern of Medicare covering non-poor population. Indeed, the link between no insurance through Medicare and the lowest retirement income confirms this pattern. Moreover, Medicare had no significant associations with middle retirement income, which linked to no private health insurance indicating this group is not covered by public nor private options. However, associations of Medicare with younger people suggest its economic contribution by covering disabled or severely ill unable to work (end stage renal disease, and Lou Gehrig's disease).[25]

The broadest coverage of 55.7% of U.S. population in 2013[26] by insurance through employer that is inherently income-dependent and stands alone in our analyses, reinforces the high relevant link of income and health insurance reproducing access barriers of an unequal access to income.

However, in the entire network, the most relevant variables suggest an option towards universal health coverage more effective for people's well-being. Highly connected to several social resources, including broader profiles of people, with a moderate but not excluding emphasis on underprivileged, the variable of no insurance through Medicare represents a potential target to better spread any intervention. The variable of insurance through Medicare complements the former variable by covering the remaining people's profiles of elderly and disabled, and supports the alternative of a more homogeneous health coverage (perhaps federal-oriented), equally available for the full population, less directly mediated by income, and even equitably progressive to meet people's health needs to strength human security instead of market sustainability.

**CONCLUSIONS**

Health care coverage is a highly-valued social resource part of different assets needed for people's well-being. However, aligned to the public health body of knowledge, our results support and explain the intermediate relevance of health coverage through insurance.

Health care coverage is just a part of well-being and of the related integral health. However, if health coverage is driven by insurance, health coverage reproduces income distribution. In this way, the role of health coverage in well-being fades even more when is strongly linked to individual`s income, if income distribution is associated to social inequality.

Therefore, even preserving their intermediary role, implementing health coverage through less income-oriented options, accordingly extended to the whole population, would better contribute to population's well-being and human development.

Beyond well-being and human development, the way of implementing the operation of any type of health coverage answers the remaining question of its impact on health results. Therefore, assessing individual health related quality of life in by similar systemic approach requires additional modeling efforts of other specific health data sources already available.

# REFERENCES

1    UN. Transforming our world: the 2030 agenda for Sustainable Development. United Nations General Assembly. Seventh session. Draft resolution referred to the United Nations summit for the adoption of the post-2015 development agenda by the General Assembly at its sixty-ninth session, 2015.

2    UN. Sustainable Development Goals - United Nations Department of Economic and Social Affairs. Sustain. Dev. Knowl. Platf. 2015. https://sustainabledevelopment.un.org/?menu=1300 (accessed Oct 15, 2015).

3    WHO. Basic Documents- Constitution of WHO-World Health Organization, 48th. edn. Geneva, Switzerland.: World Health Organization 2014, 2014.

4    UN. Millenium Development Goals. News Millenium Dev. Goals. 2015. http://www.un.org/millenniumgoals/ (accessed Oct 15, 2015).

5    Schmidt H, Gostin LO, Emanuel EJ. Public health, universal health coverage, and Sustainable Development Goals: can they coexist? *Lancet* 2015; **386**: 928–30.

6    Solar O I. A conceptual framework for action on the social determinants of Health. Social Determinants of health discussion Paper 2. Geneva: World Health Organization, 2010.

7    The World Bank Group. Health systems - Financing. World Bank. 2011. http://web.worldbank.org/WBSITE/EXTERNAL/TOPICS/EXTHEALTHNUTRITIONANDPOPULATION/EXTHSD/0,,contentMDK:22523961~menuPK:6485082~pagePK:148956~piPK:216618~theSitePK:376793,00.html (accessed Jan 1, 2016).

8    Faguet GB. The Affordable Care Act. A Missed Opportunity, A Better Way Forward. New York, U.S.: Algora Publishing, 2013.

9    Helmichen L, Kaestner R, Lo Sasso A. Advances in Health Economics and Health Services Researh. Volume 19. Beyond Health Insurance: Public Policy to Improve Health. JAI Press - Emerald Group Publishing, 2008.

10   Figueras J, McKee M. Health Systems, Health, Wealth And Societal Well-Being: Assessing the case for investing in health systems. McGraw Hill/ Open University Press, 2011.

11   Cifuentes MP, Doogan NJ, Fernandez SA, Seiber EE. Revealing the role of factors shaping Americans' objective well-being: A systems science approach with network analysis. *J Policy Model* 2016; **In press,** . DOI:doi:10.1016/j.jpolmod.2016.03.008.

12   Morrisey MA. Health Insurance in the United States. In: Dionne G, ed. Handbook of Insurance. New York, EU.: Springer Science+Business Media, 2013: 205–30.

13   Ruggles S, Genadek K, Goeken R, Grover J, Sobek M. Integrated Public Use Microdata Series: Version 6.0 [Machine-readable database]. Minneapolis: University of Minnesota, 2015. 2015.

14   U.S. Census Bureau. American Community Survey (ACS). United States Census Bur. 2015. https://www.census.gov/programs-surveys/acs/about.html (accessed Feb 2, 2015).

15   SHADAC. 'Defining Family' for Studies of Health Insurance Coverage. 2012; : 1–6.

16   Kolaczyk ED. Statistical Analysis of Network Data. Springer, 2010.

17   De Nooy W, Mrvar A, Batagelj V. Exploratory Social Network Analisis with Pajek. Cambridge University Press, 2005.

18   Batagelj V, Doreian P, Ferligoj AA, Kejzar N. Understanding Large Temporal Networks and Spatial Networks. Exploration, Pattern Searching, Visualization and Network Evolution. John Wiley and Sons, 2014.

19   Batagelj V, Mrvar A. Pajek Pajek-XXL Reference Manual. Ljubljana, Slovenia, 2014.

20   Batagelj V, Mrvar A. Pajek64.4.04 Software. 2014.

21   SHADAC. Using SHADAC Health Insurance Unit (HIU) and Federal Poverty Guideline (FPG) Microdata Variables. 2013; : 1–5.

22   CMCS. Eligibility. Medicaid.gov. 2016. https://www.medicaid.gov/affordablecareact/provisions/eligibility.html.

23   DeMichele T, Bastian C, Mullen P. Federal Poverty Level Guidelines. Obamacarefacts.com. 2015. http://obamacarefacts.com/federal-poverty-level/ (accessed Feb 2, 2016).

24   Morrisey MA. Health Insurance. Chicago, Illinois - Washington, DC.: Health Administration Press - AUPHA Press, 2008.

25   HHS. Who's eligible for Medicare? HHS.gov U.S. Dep. Heal. Hum. Serv. 2014. http://www.hhs.gov/answers/medicare-and-medicaid/who-is-elibible-for-medicare/index.html (accessed Jan 1, 2016).

26   Smith JC, Medalia C. Health Insurance Coverage in the United States: 2014. Washington, DC., 2015.